\documentclass[12pt]{article}

\begin{document}
\begin{center}
\textbf{A\ POTENTIAL\ OF\ INCOHERENT ATTRACTION BETWEEN\ MULTIDIMENSIONAL\
SOLITONS}

Andrey Maimistov$^a$, Boris Malomed$^b$\footnote{%
electronic address malomed@eng.tau.ac.il}, and Anton Desyatnikov$^a$

\vspace{1.0cm}$^a$Moscow Engineering Physics Institute, Kashirskoye sh., 31,
Moscow, 115409, Russia

$^b$Department of Interdisciplinary Studies, Faculty of Engineering, Tel
Aviv University, Tel Aviv 69978, Israel
\end{center}

\newpage

\begin{center}
\textbf{ABSTRACT}
\end{center}

We obtain analytical expressions for an effective potential of interaction
between two- and three-dimensional (2D and 3D) solitons (including the case
of 2D vortex solitons) belonging to two different modes which are
incoherently coupled by cross-phase modulation. The derivation is based on
calculation of the interaction term in the full Hamiltonian of the system.
An essential peculiarity is that, in the 3D case, as well as in the case of
2D solitons with unequal masses, the main contribution to the interaction
potential originates from a vicinity of one or both solitons, similarly to
what was recently found in the 2D and 3D single-mode systems, while in the
case of identical 2D solitons, the dominating area covers all the space
between the solitons. Unlike the single-mode systems, \emph{stable} bound
states of mutually orbiting solitons are possible in the bimodal system.

\newpage

\section{INTRODUCTION}

Interaction between self-focussing cylindrical beams (spatial solitons) in
bulk nonlinear media is a problem of obvious interest both by itself and for
applications. This interaction was studied experimentally and by means of
numerical simulations in photorefractive media \cite{Photo1,Photo2}, and was
simulated in detail in an isotropic model with the
second-harmonic-generating (SHG) nonlinearity \cite{SKB}. In the latter
model, it was demonstrated that the spiraling bound state of two cylindrical
beams is unstable. A general analytical expression for a potential of
interaction between far separated two- and three-dimensional (2D and 3D)
solitons was very recently derived in \cite{me}. The potential also predicts
an instability of the orbiting bound state of two solitons.

A very convenient model for the study of the multidimensional solitons and
their interactions is the \emph{cubic-quintic} nonlinear Schr\"odinger
(CQNLS) equation, in which the cubic nonlinearity is self-focusing, giving
rise to the beams (2D solitons) or ``light bullets'' \cite{Yaron} (3D
solitons), while the quintic term is self-defocusing, precluding the wave
collapse in the model: 
\begin{equation}
iu_t+\nabla ^2u+|u|^2u-g|u|^4u=0.  \label{CQ}
\end{equation}
The coefficients in (\ref{CQ}), except for $g$, can be set equal 
to $1$ by means of 
scale transformations, whereas $g$ is the quintic-to-cubic nonlinear
susceptibilities ratio. In the application to nonlinear
optical media, the temporal variable $t$ in (\ref{CQ}) must be replaced by
the propagation distance $z$, while the role of the third transverse
variable is played by the ``reduced time'', $t-z/V_{\mathrm{gr}}$, $V_{%
\mathrm{gr}}$ being the mean group velocity of the carrier wave (this
implies that the temporal dispersion in the medium is anomalous) \cite{Yaron}%
. Finally, the Hamiltonian of this model is 
\begin{equation}
H_u=\int \left( \left| \nabla u\right| ^2-\frac 12|u|^4+\frac g3|u|^6\right)
d\mathbf{r}.  \label{H_0}
\end{equation}

\emph{Vortex} beams, with the vorticity (``spin'') $s=1$, and interactions
between them, described by Eq. (\ref{CQ}), were simulated by
Quiroga-Teixeiro and Michinel \cite{Manolo}. A remarkable result is the
numerically discovered robustness of the vortex beams (which were found to
be strongly unstable against azimuthal perturbations in the SHG model \cite
{azimuth}). Note that the model (\ref{CQ}) is not merely the simplest one
that gives rise to stable multidimensional solitons: according to
experimental data \cite{PTS}, the combination of the focusing cubic and
defocusing quintic terms adequately represents the nonlinear optical
properties of some real materials.

The effective potential of the intersoliton interaction derived in \cite{me}
applies to a wide class of models, including Eq. (\ref{CQ}). However, it
does not apply to bimodal systems including \emph{two} equations with \emph{
in}coherent nonlinear coupling between them (mediated by the cross-phase
modulation in nonlinear optical media), in the case when the two solitons
(beams) belong to different modes. The simplest bimodal generalization of
the model based on the Hamiltonian (\ref{H_0}) is furnished by the
Hamiltonian $H=H_u+H_v+H_{\mathrm{int}}$, where $H_v$ is the same expression
as (\ref{H_0}) with the field $u$ substituted by another field $v$, and the
interaction part of the Hamiltonian is 
\begin{equation}
H_{\mathrm{int}}=\int \left[ -\beta |u|^2|v|^2+\alpha \left(
|u|^4|v|^2+|u|^2|v|^4\right) \right] d\mathbf{r},  \label{H_int}
\end{equation}
with, generally speaking, arbitrary positive constants $\alpha $ and 
$\beta $. The full Hamiltonian of the bimodal system gives 
rise to the equations 
\begin{eqnarray}
iu_t+\nabla ^2u+\left( |u|^2+\beta \left| v\right| ^2\right) u-\left(
g|u|^4+2\alpha |u|^2|v|^2+\alpha \left| v\right| ^4\right) u &=&0,  \label{u}
\\
iv_t+\nabla ^2v+\left( |v|^2+\beta \left| u\right| ^2\right) u-\left(
g|v|^4+2\alpha |u|^2|v|^2+\alpha \left| u\right| ^4\right) v &=&0.  \label{v}
\end{eqnarray}
Commonly known examples of optical bimodal systems are provided by two
orthogonal polarizations of light, or two light waves with different carrier
wavelengths \cite{Agrawal}. In the latter case, as well as in the former one
with the circular polarizations, the cubic cross-phase-modulation
coefficient is $\beta =2$. In the case of two linear polarizations, $\beta
=2/3$ (and the usual assumption is to drop additional four-wave mixing
terms). The constant $\alpha $ is left here to be arbitrary, but, in the
most interesting cases, the second term in (\ref{H_int}) will only produce a
small correction to the effective interaction potential.

As well as in the case of the single-mode system, the interaction between
solitons in different modes depends on the separation between them, but,
unlike the single-mode case, it is not sensitive to a phase difference
between the solitons, hence the interaction is expected to be \emph{simpler}
than inside the same mode. The objective of this work is to find an
effective potential of
the interaction between 2D and 3D solitons in the bimodal
system, including the cases when the interacting solitons are both identical
and different (the interaction between 2D solitons with different
vorticities is also included). The interaction between identical 2D beams
was recently considered in \cite{Moscow}, but using an artificially
introduced Gaussian ansatz for the soliton. As well as it was done in \cite
{me} for the single-mode system, in this work we find the interaction
potential in
a general analytical form. However, the derivation is essentially
different from that developed in \cite{me}; in particular, the derivation
proves to be very different for the 2D and 3D cases, while in the
single-mode system these two cases were very similar. In section 2, we
derive the potential for the interaction between 2D solitons (spatial beams)
with unequal masses. We explicitly consider two limit cases, when the masses
of the interacting solitons are very different or nearly equal. The latter
case demonstrates a singularity in the limit of equal masses, therefore the
interaction between identical 2D solitons should be considered separately,
which is done in section 3. The result, and the way to obtain it, turn out
to be drastically
different from the case of unequal masses: when the masses are
not equal, a dominating contribution to the effective interaction potential
is produced by a vicinity of the soliton having a larger mass (which is
similar to the situation for the single-mode system \cite{me}), while, in
the equal-mass case, a dominating area is located \emph{between} the
solitons, in contrast with the case of the single-mode system.
In section 4, the potential is derived for the 3D solitons with equal
masses. In this case, the derivation is similar to that for the 2D solitons
with the {\emph un}equal 
masses, but it gives rise to an additional logarithmic factor.

The results are summarized in section 5, where we conclude, in particular,
that the obtained potentials give rise to two bound states of the solitons
orbiting around each other, one of which is \emph{stable} (which was already
concluded in \cite{Moscow}), on the contrary to the orbiting states in the
single-mode models \cite{me} (including the SHG one \cite{SKB}), which are
all unstable. This difference, which is, obviously, very important for
applications, is due to the fact that, in the bimodal system, the
interaction potential does not depend on the phase difference between the
two solitons.

\section{THE\ INTERACTION\ BETWEEN\ DIFFERENT TWO-DIMENSIONAL\ SOLITONS}

A general 2D stationary solution to Eq. (\ref{u}), with an internal
frequency $\omega \equiv -\mu ^2$, is looked for in the form 
\begin{equation}
u_s=\exp \left( i\mu ^2t+is\theta \right) U(r),  \label{2D}
\end{equation}
where the $s$ is an integer spin (vorticity), and a real function $U(r)$
satisfies the equation $U^{\prime \prime }+r^{-1}U^{\prime
}-s^2r^{-2}+U^3-gU^5=\mu ^2U$, which can be easily solved numerically \cite
{Manolo}. A soliton solution is defined by its asymptotic form at 
$r\rightarrow \infty $, 
\begin{equation}
U(r)\approx A_s(\mu )\;r^{-1/2}\exp (-\mu r),  \label{asymp2D}
\end{equation}
where the amplitude $A_s(\mu )$ is to be found numerically. We will consider
a situation with the solitons of the form (\ref{2D}), (\ref{asymp2D}) in
each mode $u$ and $v$, that have, generally, different spins $s_1$ and $s_2$
and different frequency parameters $\mu _1$ and $\mu _2$ which determine
effective masses of the solitons. A size of the soliton can be estimated,
pursuant to Eq. (\ref{asymp2D}), as $\mu ^{-1}$. We will consider the case
when the separation between the solitons is much larger than their proper
sizes, i.e., $R\mu _{1,2}\gg 1$.

The interaction Hamiltonian (\ref{H_int}) allows one to define an effective
interaction potential for two separated solitons, approximating the
two-soliton configuration by a linear superposition of two isolated
solitons, and substituting it into (\ref{H_int}) \cite{GO}. This approach
requires actual calculation of the integrals in (\ref{H_int}), which can be
done for 2D solitons in an exact form only in exceptional cases (see, e.g., 
\cite{Step}), another drawback being that a distortion of the ``tail'' of
each soliton due to its interaction with the ``body'' of the
other one is
ignored. In the work \cite{Moscow}, the necessary integral was evaluated,
assuming a Gaussian ansatz for the isolated solitons. However, 
the corresponding
effective interaction potential, decaying $\sim \exp \left( -\mathrm{const}
\cdot R^2\right) $, was actually produced by the ansatz rather than by the
model. In fact, the potential must decay as $\exp \left( -\mathrm{const}
\cdot R\right) $, see below.

In the work \cite{me}, another approach to the calculation of the effective
potential was developed for the single-mode systems, following the way
elaborated earlier for 1D solitons in \cite{1D}. This method did not require
knowing internal structure of the soliton, and did not imply neglecting the
distortion of each soliton's tail due to its overlapping with the other
soliton. All that is necessary to know about the individual solitons for the
application of this method, is only the asymptotic amplitudes $A_s(\mu )$ in
(\ref{asymp2D}). Here, we will apply a similar method to the bimodal 
system (\ref{u}), (\ref{v}), although technical details 
will be essentially different
from those in the case of the single-mode systems.

We start, still assuming the linear superposition of the two solitons 
$u_{s_1}$ and $v_{s_2}$ (see Eqs. (\ref{2D}) and (\ref{asymp2D})) with widely
separated centers, and setting, for the definiteness, $\mu _1>\mu _2$.
Because of the exponential decay of the fields, it is obvious that a
dominant contribution to the interaction potential (\ref{H_int}) will be
produced by a vicinity of the soliton with $\mu =\mu _1$.

First, we consider the case $\mu _2\ll \mu _1$, i.e., a light soliton (beam)
interacting with a heavy one. Substituting the field $v$ for the light
soliton by the asymptotic expression (\ref{asymp2D}), and neglecting its
small variation over the size of the narrow heavy soliton, we can easily
cast the expression for the interaction potential into the form 
\begin{eqnarray}
H_{\mathrm{int}} &\approx &A_{s_2}^2(\mu _2)R^{-1}\exp \left( -2\mu
_2R\right) \left[ -\beta \int \left| u_{s_1}(\mathbf{r;}\mu _1)\right| ^2d
\mathbf{r}+\alpha \int \left| u_{s_1}(\mathbf{r;}\mu _1)\right| ^4
d\mathbf{r}\right]  \nonumber \\
&\equiv &\left( -\beta m_1+\alpha \widetilde{m}_1\right) A_{s_2}^2(\mu
_2)R^{-1}\exp \left( -2\mu _2R\right) ,  \label{light}
\end{eqnarray}
where $m_1\equiv \int \left| u_{s_1}(\mathbf{r;}\mu _1)\right| ^2d\mathbf{r}$
and $\widetilde{m}_1\equiv \int \left| u_{s_1}(\mathbf{r;}\mu _1)\right| ^4d
\mathbf{r}$ are two integral characteristics of the heavy soliton, $m_1$
being, in fact, its effective mass. Thus, both attraction and repulsion
between the light and heavy solitons may take place, depending on the sign
in front of the expression (\ref{light}).

Another interesting case is $\mu _1-\mu _2\equiv \Delta \mu \ll $ $\mu
_2\equiv \mu $ (i.e., the interaction between nearly identical solitons,
provided that $s_1=s_2$). In this case, following \cite{me}, we assume that,
in terms of the polar coordinates $(r,\theta )$ with the origin at the
center of the heavier (first) soliton, a main contribution to $H_{\mathrm{
int}}$ comes from the distances $\mu ^{-1}\ll r\ll R$, where \emph{both}
solitons may be approximated by the asymptotic expressions (\ref{asymp2D}).
Then, it is straightforward to obtain the following expression corresponding
to the first term in (\ref{H_int}) (cf. the corresponding expressions and
Fig. 1 in \cite{me}): 
\[
U_{\mathrm{int}}\approx -\beta A_{s_1}^2(\mu )A_{s_2}^2(\mu
)R^{-1}\int_0^\infty rdr\int_0^{2\pi }d\theta \;r^{-1} 
\]
\begin{equation}
\exp \left[ -2(\mu +\Delta \mu )r-2\mu \sqrt{(R+r\cos \theta )^2+r^2\sin
^2\theta }\right] .  \label{Delta_mu}
\end{equation}
Here, the small difference $\Delta \mu $, and the difference between $R$ and
the exact distance from an integration point $(r,\theta )$ to the center of
the second (lighter) soliton, $\sqrt{(R+r\cos \theta )^2+r^2\sin ^2\theta }$
, are neglected everywhere, except for the argument of the exponential
function. Making use of the expansion 
\begin{equation}
\sqrt{(R+r\cos \theta )^2+r^2\sin ^2\theta }=R+r\cos \theta +\mathcal{...},
\label{expansion}
\end{equation}
and of the formula $\int_0^{2\pi }\exp \left( -2\mu r\cos \theta \right)
d\theta \approx \sqrt{\pi /\mu r}\exp \left( 2\mu r\right) $, valid 
for $\mu~r$~$\gg~1$, we can simplify the integral (\ref{Delta_mu}) 
to the form 
\begin{eqnarray}
H_{\mathrm{int}} &\approx &-\beta A_{s_1}^2(\mu )A_{s_2}^2(\mu )\sqrt{\pi
/\mu }R^{-1}\exp (-2\mu R)\int_0^\infty r^{-1/2}\exp \left[ -2(\Delta \mu
)r\right] dr  \nonumber \\
&\equiv &-\pi \beta \left( 2\mu \Delta \mu \right) ^{-1/2}A_{s_1}^2(\mu
)A_{s_2}^2(\mu )R^{-1}\exp (-2\mu R).  \label{Delta_mu_final}
\end{eqnarray}
For small $\Delta \mu $, the integral in (\ref{Delta_mu_final}) is dominated
by a contribution from the region $r\sim 1/\Delta \mu \gg \mu ^{-1}$, which
justifies the use of the asymptotic approximation (\ref{asymp2D}) for the
field $u(r)$. A contribution from the second term in the full expression 
(\ref{H_int}), evaluated in the same approximation, demonstrates the same
dependence on the separation $R$, but without the large multiplier $(\Delta
\mu )^{-1/2}$, therefore this only a small correction to (\ref
{Delta_mu_final}).

\section{ATTRACTION BETWEEN IDENTICAL TWO-DIMENSIONAL SOLITONS}

The expression (\ref{Delta_mu_final}) diverges in the most interesting case 
$\Delta \mu =0$, which corresponds to the interaction between identical
solitons (provided that $s_1=s_2\equiv s$). The divergence suggests that a
region dominating the interaction potential is not that around the soliton,
as was the case both in the previous section for the case of $\Delta \mu
\neq 0$, and, for the identical solitons, in the single-mode system \cite{me},
but a wider region \emph{between} the two solitons. To calculate $U_{
\mathrm{int}}$ in this case, we use the Cartesian coordinates $(x,y)$
defined so that the centers of the two solitons are placed at the points 
$(\pm R/2,0)$. Then, using once again the asymptotic expressions (\ref
{asymp2D}) for both $u$- and $v$-solitons with $\mu _1=\mu _2\equiv \mu $,
the interaction potential corresponding to the first term in Eq. (\ref{H_int}
) is given, after obvious transformations, by 
\[
H_{\mathrm{int}}=-\beta A_s^4\int \int d\xi d\eta \;\left[ \left( \xi
^2+\eta ^2+1/4\right) ^2-\xi ^2\right] ^{-1/2} 
\]
\begin{equation}
\exp \left[ -2\mu R\left( \sqrt{\left( \xi +1/2\right) ^2+\eta ^2}+\sqrt{
\left( \xi -1/2\right) ^2+\eta ^2}\right) \right] ,  \label{ident}
\end{equation}
where $\xi \equiv x/R$, and $\eta \equiv y/R$. Because the parameter $\mu R$
is large according to the underlying assumption, the integral (\ref{ident})
is dominated by a contribution from a vicinity of points where the argument
of the exponential function has a maximum. An elementary analysis shows that
the maximum is attained not at isolated points, but rather at the 
whole segment 
$|\xi |<1/2$, $\eta =0$. Expanding the integral in small $\eta ^2$ in a
vicinity of this segment, we approximate Eq. (\ref{ident}) by an integral
that can be easily calculated: 
\[
H_{\mathrm{int}}=-\beta A_s^4\int_{-1/2}^{+1/2}d\xi \left( \frac 14-\xi
^2\right) ^{-1}\int_{-\infty }^{+\infty }d\eta \exp \left[ -2\mu R\left(
1+\frac 12\frac{\eta ^2}{1/4-\xi ^2}\right) \right] 
\]
\begin{equation}
=-\beta A_s^4\sqrt{\frac{\pi ^3}{\mu R}}\exp (-2\mu R).  \label{ident_final}
\end{equation}
Comparing this result with that (\ref{Delta_mu_final}) for the solitons with
different masses, we conclude that the divergence in the latter expression
at $\Delta \mu \rightarrow 0$ implies replacement of the pre-exponential
factor $R^{-1}$ by a larger one, $R^{-1/2}$. Evaluating the second term in 
(\ref{H_int}) in the same approximation, we conclude that it yields an
expression differing from (\ref{ident_final}) just by the factor $R^{-1}$
instead of $R^{-1/2}$, i.e., a small correction to (\ref{ident_final}). Note
that, unlike the interaction between heavy and light solitons, which may
have either sign, the nearly identical or identical solitons always attract
each other, cf. Eqs. (\ref{light}), (\ref{Delta_mu_final}), and (\ref
{ident_final}).

\section{ATTRACTION\ BETWEEN\ THREE-DIMENSIONAL\ SOLITONS}

In the 3D case, we consider only the solitons without the internal ``spin'',
i.e., with $s=0$. The 3D soliton has the form $u=\exp (i\mu ^2t)\;a(r)$,
with the asymptotic form $a(r)\approx Ar^{-1}\exp (-\mu r)$ at $r\rightarrow
\infty $, cf. Eqs. (\ref{2D}) and (\ref{asymp2D}). Substitution of this
asymptotic expression for the fields $u$ and $v$ into the integral in the
first term of Eq. (\ref{H_int}) around each soliton (cf. Eq. (\ref{Delta_mu}
)) and using the expansion (\ref{expansion}) yield 
\[
H_{\mathrm{int}}=-4\pi \beta A^4R^{-2}\int_0^\infty r^2dr\int_0^\pi \sin
\theta \;d\theta \;r^{-2}\exp \left[ -2\mu r-2\mu \left( R+r\cos \theta
\right) \right] , 
\]
where an extra factor $2$ takes into regard the fact that, in the case of
identical solitons, one has equal contributions from the vicinity of both
solitons. After elementary integration over $\theta $, we arrive at an
expression containing a formal logarithmic singularity, 
\begin{equation}
H_{\mathrm{int}}=-2\pi \beta A^4\mu ^{-1}R^{-2}\exp \left( -2\mu R\right)
\int_0^\infty r^{-1}dr.  \label{3D}
\end{equation}
Actually, the lower and upper limits of the integration are, respectively, 
$\sim ~\mu ^{-1}$ and $\sim ~R$, so that, with the 
\emph{logarithmic accuracy},
the final expression for the effective interaction potential in the 3D
case becomes (cf. (\ref{ident_final})) 
\begin{equation}
H_{\mathrm{int}}=-2\pi \beta A^4\mu ^{-1}R^{-2}\exp \left( -2\mu R\right)
\ln (\mu R).  \label{3Dfinal}
\end{equation}
As for the contribution from the second term in (\ref{H_int}), it has the
same dependence on $R$ as (\ref{3Dfinal}), but without the large logarithmic
factor, so that in this case too, it is a small correction only.

The above analysis was done for identical solitons. If the solitons have a
small mass difference, corresponding to a small difference $\Delta \mu $,
the interaction potential is given by essentially the same expression (\ref
{3Dfinal}), except for a factor of $2$, which is absent if $\Delta \mu \cdot
R\;_{\sim }^{>}\;1$, when the calculation of $H_{\mathrm{int}}$ is dominated
by a vicinity of one soliton only.

\section{CONCLUSION}

In this work, we have derived analytical expressions for an effective
potential of interaction between two- and three-dimensional solitons
(including the case of the two-dimensional vortex solitons) belonging to two
different modes which are incoherently coupled through cross-phase
modulation in models of media with the self-focusing cubic and
self-defocusing quintic nonlinearities. The derivation was based on the
calculation of the interaction term in the full Hamiltonian of the system.
An essential peculiarity is that, in the 3D case, as well as in the case of
2D solitons with unequal masses, the main contribution to the interaction
potential originates from a vicinity of one or both solitons, similarly to
what was recently found in the 2D and 3D single-mode systems \cite{me},
while in the case of identical 2D solitons, the dominating area covers all
the space between the solitons. Except for the case of the interaction
between light and heavy solitons, which may have either sign, the solitons
always attract each other.

The attraction between the solitons may give rise to their orbiting bound
states in the 2D and 3D cases (in the latter case, it is assumed that the
two solitons move in one plane). Orbiting of incoherently interacting 2D
solitons was experimentally observed in a photorefractive medium \cite
{Photo1}. Numerical simulations and analytical arguments presented in \cite
{SKB} and \cite{me} demonstrate that the orbiting bound states of the 2D
solitons in the single-mode systems, including the SHG model, are unstable.
However, it was recently pointed out in \cite{Moscow} that they might be
stable in the bimodal system. Indeed, for the orbiting state, the
interaction potential (\ref{Delta_mu_final}), (\ref{ident_final}), or (\ref
{3Dfinal}) must be supplemented by the centrifugal energy $E_{\mathrm{cf}
}=\left( M^2/2m\right) R^{-2}$, where $M$ is the angular momentum of the
soliton pair, and $m$ is the soliton's effective mass. A straightforward
consideration of the net potential, $H_{\mathrm{int}}+E_{\mathrm{cf}}$,
demonstrates that it may have two stationary points, the one corresponding
to a smaller value of $R$ being a potential \emph{minimum} that gives rise
to a stable orbiting state. The instability of similar states in the
single-mode systems is due to the fact that, in those systems, the
interaction potential also depends on the phase difference between the
solitons, the effective mass corresponding to the phase-difference degree of
freedom being \emph{negative} \cite{Seva}. This, eventually, made the
existence of stable stationary points of the effective interaction potential
impossible.

This work has been supported by INTAS under grant No: 96-0339

\end{document}